\documentclass[11pt]{article}

\usepackage{amsthm, amsfonts}
\usepackage[margin=2.5cm]{geometry}
\usepackage{amssymb}
\usepackage[]{amsmath}
\usepackage{fancyhdr}
\usepackage[cp1250]{inputenc}

\newtheorem{thm}{Theorem}[section]
\newtheorem{rem}{Remark}[section]

\usepackage{hyperref}
\hypersetup{
    colorlinks=true,
    linkcolor=blue,
    filecolor=magenta,      
    urlcolor=cyan,
}

\title{Optimal evolution models for quantum tomography}
\author{\Large Artur Czerwi\'nski\textsuperscript{1,2}\\
\\
1. Center for Theoretical Physics of the Polish Academy of Sciences\\
Al. Lotnikow 32/46, 02-668 Warsaw, Poland\\
\\
2. Institute of Physics, Faculty of Physics, Astronomy and Informatics,\\
Nicolaus Copernicus University,\\
Grudziadzka 5, 87-100 Torun, Poland\\
Corresponding author's e-mail:  aczerwin@fizyka.umk.pl
}
\begin{document}
\maketitle

\begin{abstract}
The research presented in this article concerns the stroboscopic approach to quantum tomography, which is an area of science where quantum Physics and linear algebra overlap. In this article we introduce the algebraic structure of the parametric-dependent quantum channels for 2-level and 3-level systems such that the generator of evolution corresponding with the Kraus operators has no degenerate eigenvalues. In such cases the index of cyclicity of the generator is equal $1$, which physically means that there exists one observable the measurement of which performed sufficient number of times at distinct instants provides enough data to reconstruct the initial density matrix and, consequently, the trajectory of the state. Necessary conditions for the parameters and relations between them are introduced. The results presented in this paper seem to have considerable potential applications in experiments due to the fact that one can perform quantum tomography by conducting only one kind of measurement. Therefore, the analyzed evolution models can be considered optimal in the context of quantum tomography. Finally, we also introduce some remarks concerning optimal evolution models in case of $n-$dimensional Hilbert space.
\end{abstract}


\maketitle

\section{Introduction}\label{sec1}
For the first time a quantum tomography problem was formulated in 1933 when Pauli was considering whether the quantum wavefunction of a system is uniquely determined by its position and momentum probability distributions \cite{pauli80,reich44}. Currently it is commonly known that in general Pauli's problem is not uniquely solvable for any wavefunction \cite{reich44,corbett06}. Since 1933 there have been many proposals to tackle the problem of quantum wavefunction reconstruction such as Gerchberg-Saxton algorithm \cite{gerchberg72}. Nowadays the term quantum tomography refers to all methods which aim to reconstruct the quantum representation of a physical system.

According to one of the most fundamental assumption of quantum mechanics the density matrix $\rho$ contains all the information about the analyzed quantum system. The problem of quantum tomography, i.e. reconstructing the initial density operator $\rho(0)$ on the basis of experimental data, has been receiving much attention in recent years as the ability to control and manipulate quantum states can be transferred to other areas of Physics, such as quantum communication and computing. A successful model of tomography can be created only if it is assumed that a set of identically prepared copies of the quantum state is available. Thus each quantum system is measured only once, which makes irrelevant the problem of changes in quantum state due to measurement. Physical motivation for research into quantum tomography is obvious -- a successful tomography model can be used in quantum communication or quantum optics to verify whether the system is prepared in the desirable state. What makes this issue even more interesting is the fact that quantum tomography is the area where quantum physics and linear algebra meet. It would not have been possible to determine optimal criteria for state reconstruction if we had not analyzed the problem by means of algebraic methods.

If one considers a static quantum tomography problem for a system associated with Hilbert space $\mathcal{H}$ such that $dim \mathcal{H} = n$, it is necessary to measure $n^2 - 1$ different physical quantities (cf. \cite{altepeter04,alicki87,genki03}). The obvious problem that arises in this approach is connected with the possibility to implement the theoretical model of tomography in an experiment. In the static model the number of observables increases quadratically with the dimension of the Hilbert space. For example, in case of $dim \mathcal{H} = 4$ one would have to find mean values of 15 Hermitian operators in order to determine the density matrix. Mathematically, it is not difficult to propose such number of Hermitian operators, but in laboratory reality it appears impossible to find 15 physical quantities that can be measured. Thus the practical aspects of this approach seem rather questionable.

Therefore, in this paper we follow the stroboscopic approach to quantum tomography, which was founded in 1983 in the paper \cite{jam83} and then developed in many papers such as \cite{jam00,jam04,jam12}. In this approach we assume to have a set of observables $\{Q_i\}_{i=1} ^r $ (where $r< n^2 -1$) and each of them can be measured at different time instants. Every measurement provides a result (denoted by $m_i (t_j)$) which can be expressed as $m_i (t_j) = Tr(Q_i \rho(t_j))$. We say that the quantum system is $(Q_1, \dots, Q_r)$-reconstructible on the interval $[0,T]$ if there exists a sequence of time instants $0\leq t_1 < t_2< \dots < t_p \leq T$ such that the initial density matrix $\rho(0)$ can be calculated from the measurement results $m_i(t_j)$ where $i=1,\dots, r$ and $j=1, \dots, p$. Because in the stroboscopic tomography the measurements are performed at different time instants, we need to assume that the knowledge about the evolution of the system is available, e.g. the Kossakowski-Lindblad master equation \cite{gorini76,lindblad76} is given or, equivalently, the collection of Kraus operators. The assumption about the character of evolution enables us to reconstruct not only the initial state, but also the complete trajectory of the quantum state.

The main questions that arise in this approach concern: for a given master equation what is the minimal number of observables? What are properties of the observables? What is the minimal number of time instant and how to choose them? The general conditions for observability have been determined and can be found along with the proofs in the papers \cite{jam83,jam00,jam04}.

From all the questions that relate to the stroboscopic tomography in this article we mainly focus on the problem of the minimal number of observables required to reconstruct the quantum state. Let us revise a general theorem on the minimal number of observables.
\begin{thm}\label{The1}
For a quantum system which evolution is given by the Kossakowski-Lindblad master equation of the form
\begin{equation}\label{eq:kossakowski}
\frac{ d \rho}{d t } = \mathbb{L} [\rho], 
\end{equation}
where the operator $\mathbb{L}$ is called the generator of evolution, there exists a number (denoted by $\eta$) which expresses the minimal number of observables required to reconstruct the density matrix and is called \textit{the index of cyclicity}. The index of cyclicity can be computed from the formula \cite{jam83}
\begin{equation}
\eta := \max \limits_{\lambda \in \sigma (\mathbb{L})} \{ dim Ker (\mathbb{L} - \lambda \mathbb{I})\},
\end{equation}
where by $\sigma (\mathbb{L})$ we denote the spectrum of the generator $\mathbb{L}$.
\end{thm}
One can notice that the index of cyclicity can be understood as the greatest geometric multiplicity from all eigenvalues of $\mathbb{L}$.

One can observe that the value of the index of cyclicity depends only on algebraic properties of the generator of evolution, whereas its interpretation is relevant from physical point of view -- the lower the value is, the more beneficial it is to employ the stroboscopic approach rather than static tomography. Thus, one can agree that the notion of the index of cyclicity connects in a perfect way quantum Physics with linear algebra.

One should also bear in mind that the most general form of $\mathbb{L}[\rho]$ can be expressed as \cite{gorini76,lindblad76}
\begin{equation}\label{eq:generator}
\mathbb{L} [\rho] = - i [H,\rho] +  \sum_{i=1} ^{n^2 -1} \gamma_i \left ( V_i \rho V_i ^* - \frac{1}{2} \{V_i ^* V_i, \rho\} \right ),
\end{equation}
where $\gamma_i \geq 0$, $H \in B_* (\mathcal{H})$ and $V_i \in B(\mathcal{H})$. $B_* (\mathcal{H})$ denotes the linear space of all self-adjoint operators in $\mathcal{H}$ and $B(\mathcal{H})$ the linear space of all bounded operators in $\mathcal{H}$. The equation \ref{eq:kossakowski} with the generator given by \ref{eq:generator} is the most general type of Markovian and time-homogeneous master equation which preserves trace and positivity. The above structure is commonly known and, therefore, it does not require any special attention. Nevertheless, the generator of evolution can be equivalently presented in the explicit matrix form which is obtained by employing the idea of vectorization. To transform the generator of evolution given by \ref{eq:generator} we apply the equation that connects the standard matrix product with the Kronecker product \cite{henderson81}, i.e.
\begin{equation}\label{eq:henderson}
vec(XYZ) = (Z^T \otimes X) vec Y,
\end{equation}
where $X,Y,Z$ are matrices selected in such a way that the matrix product poduct $XYZ$ is computable. 

Taking into account the relation \ref{eq:henderson} one transforms the generator of evolution given originally by \ref{eq:generator} into the matrix form
\begin{eqnarray}\label{eq:generatormatrix}
 \mathbb{L} = -i (H \otimes \mathbb{I}_n - \mathbb{I}_n \otimes H) + \sum_{i=1}^{n^2-1}  \gamma_i \left ( \overline{V}_i \otimes V_i - \frac{1}{2} \mathbb{I}_n \otimes V_i ^* V_i - \frac{1}{2} V_i ^T \overline{V}_i \otimes \mathbb{I}_n \right ),
\end{eqnarray}
where $\overline{V}_i$ refers to the complex conjugate of the operator $V_i$. The explicit matrix form of the generator of evolution is useful in the context of the stroboscopic tomography as it allows to determine properties of the generator such as its spectrum. Due to the fact that the generator of evolution according to \ref{eq:generator} consists of commutators and anticommutators it is often necessary to refer to mathematical papers devoted to properties of such operators \cite{pacyfik71,marcus64}.

From the theorem \ref{The1} one can calculate $\eta$, which is the minimal number of observables required for quantum tomography, but in order to determine the algebraic structure of these observables one needs to follow another theorem \cite{jam83,jam00}. 
\begin{thm}\label{The2}
The quantum system is $(Q_1,...Q_{\eta})$-reconstructible if and only if the operators $\{Q_1, \dots, Q_{\eta}\}$ satisfy the condition 
\begin{equation}
\bigoplus \limits_{i=0}^{\eta} K_\mu (\mathbb{L},Q_i) = B_*(\mathcal{H}),
\end{equation}
where $\bigoplus$ denotes the Minkowski sum of subspaces, $\mu$ is the degree of the minimal polynomial of $\mathbb{L}$ and $K_\mu (\mathbb{L}, Q_i)$ denotes Krylov subspace, which is defined as
\begin{equation}
 K_\mu (\mathbb{L}, Q_i) := Span \{ Q_i, \mathbb{L}^* [Q_i], (\mathbb{L}^*)^2 [Q_i], ...,(\mathbb{L}^*)^{\mu-1} [Q_i] \}.
\end{equation}
\end{thm}
\begin{rem}
In the theorem \ref{The2} we denote by $K_{\mu} (\mathbb{L},Q_0)$ an identity matrix of the appropriate dimension.
\end{rem}

The necessity to consider Krylov subspaces appears naturally in the context of stroboscopic tomography, however the theorem \ref{The2} was formulated differently in the paper \cite{jam83}, which initiated the stroboscopic tomography. The most recent formulation of this theorem can be found in \cite{jam12} as well as in \cite{czerwin15} along with a thorough explanation of its origins. The main idea behind this theorem is the polynomial representation of the quantum semigroup being the solution of the evolution equation \ref{eq:kossakowski}:
\begin{equation}\label{exponent}
\rho (t) = exp(\mathbb{L}t) \rho(0) = \sum_{k=0}^{\mu-1} \alpha_k(t) \mathbb{L}^k [\rho(0)],
\end{equation}
where $\mu$ is the degree of the minimal polynomial of $\mathbb{L}$ and $\alpha_k (t)$ are certain functions that can be calculated from a set of differential equations \cite{jam04}. The representation \ref{exponent} allows one to write the formula for the result of measurement of the observable $Q_i$ in time instant $t_j$:
\begin{equation}\label{measure}
m_i (t_j) = \sum_{k=0}^{\mu-1} \alpha_k (t_j) Tr\left( (\mathbb{L}^*)^k [Q_i] \rho(0)\right),
\end{equation}
where $\mathbb{L}^*$ denotes the dual operator to $\mathbb{L}$. From the equation \ref{measure} one can observe that if the observable $Q_i$ is measured at $\mu$ distinct time instants, then one obtains a set of $\mu$ equations from which one can calculate the projections $Tr\left( (\mathbb{L}^*)^k [Q_i] \rho(0)\right) \equiv \langle (\mathbb{L}^*)^k [Q_i] | \rho(0) \rangle$ where $k=0,\dots, \mu-1$. Then naturally $\rho(0)$ can be reconstructed iff the operators $(\mathbb{L}^*)^k [Q_i] $ for $k=0,\dots, \mu-1$ and $i=1, \dots, \eta$ span the space to which $\rho(0)$ belongs, which is stated in the theorem \ref{The2}. The theorem \ref{The2} can be put into other words -- it is necessary for quantum tomography that the operators $(\mathbb{L}^*)^k [Q_i] $ for $k=0,\dots, \mu-1$ and $i=1, \dots, \eta$ constitute a spanning set. Clearly, the condition in \ref{The2} corresponds with the four equivalent definitions of spanning set presented in \cite{ariano2000} -- the terminology is different but the main idea which concerns the criteria for quantum tomography is the same. The methods used in the stroboscopic tomography are also strictly connected with the \textit{fusion frames} theory, where it is commonly discussed under which conditions a complex vector can be reconstructed from modulus of inner product with frame vectors (cf. \cite{casazza14,bandeiraa14,conca15}).

From theorem \ref{The1} we see that the index of cyclicity is the most important quantity when somebody is considering the usefulness of the stroboscopic approach to quantum tomography. The lower the value of the index of cyclicity the more beneficial it is to employ the stroboscopic approach. Therefore, it appears justifiable to take interest in the generators with the index of cyclicity equal $1$, because in such cases there exists one observable the measurement of which repeated certain number of times provides sufficient data to reconstruct the initial density matrix. The generators of evolution with the index of cyclicity equal $1$ can be considered \textit{optimal evolution models} in reference to quantum tomography. Therefore, the goal of this paper is to determine the algebraic structure of optimal evolution models. Such optimal evolution models are introduced in this paper in two equivalent ways -- by completely positive and trace-preserving maps given in the Kraus forms and by generators of evolution. The results presented in this article are a generalization of the notion of \textit{one-parametric non-degenerate family of Kraus operators} which was introduced in \cite{czerwin15}.

In section \ref{sec2} of this article we introduce a three-parametric non-degenerate family of Kraus operators for 2-level systems. We also introduce the parametric-dependent observable which allows to perform quantum tomography provided it is measured at three distinct time instants. It is also shown that the one-parametric family introduced in \cite{czerwin15} is a special case of the more general three-parametric family proposed in this article. Then in section \ref{sec3} we take into consideration 3-level quantum systems, which results in introducing a six-parametric non-degenerate family of Kraus operators. In the last section we propose some general remarks concerning optimal evolution models for $n-$level quantum systems.

\section{Optimal evolution models for 2-level systems}\label{sec2}

Before introducing the main result let us make two remarks to explain the denotations.
\begin{rem}
In this section we assume that the Kraus operators which constitute a dynamical map for 2-level quantum systems are proportional to the Pauli matrices and the $2-$dimensional identity matrix. The Pauli matrices shall be denoted as $\{ \sigma_1, \sigma_2,\sigma_3\}$ and the corresponding identity matrix by $\mathbb{I}_2$. 
\end{rem}
\begin{rem}
The time dependency of Kraus operators shall be given by the decoherence function $\kappa (t)$ which can be expressed as $\kappa (t) = e^{-\gamma t}$, where $\gamma \in \mathbb{R}_+$ is a positive constant (compare with the decoherence models presented in \cite{nielsen00}).
\end{rem}
In this section we propose to generalize the idea of non-degenerate family of Kraus operators for 2-level systems to a three-parametric case, which seems the most general. We shall prove the following theorem.
\begin{thm}\label{The3}
The three-parametric family of Kraus operators $\{ K_i (t;a_1,a_2,a_3)\}_{i=0} ^3$ given by
\begin{eqnarray}
K_0 (t;a_1,a_2,a_3)= \sqrt{1-(a_1+a_2+a_3)(1 - \kappa(t))} \mathbb{I}_2\label{eq:kraus2} \\
K_i (t; a_i) = \sqrt{a_i(1-\kappa(t))} \sigma_i   \hspace{0.5cm}for\hspace{0.5cm} i = 1,2,3,\label{eq:kraus2con}
\end{eqnarray}
where $a_1,a_2,a_3$ are parameters that influence the structure of the generator of evolution, can be considered a three-parametric non-degenerate family of Kraus operators only if the parameters fulfill the following relations
\begin{eqnarray}
a_1+a_2+a_3 \leq 1  \hspace{0.5cm}and\hspace{0.5cm}  a_1,a_2,a_3 \in \mathbb{R}_+ \cup \{0\} \label{eq:condition1}\\
a_1 \neq a_2  \neq a_3  \label{eq:condition2}
\end{eqnarray}

In other words, the theorem claims that the generator of evolution corresponding with the dynamical map given by Kraus operators from \ref{eq:kraus2}-\ref{eq:kraus2con} has no degenerate eigenvalues provided the parameters $(a_1,a_2,a_3)$ fulfill the conditions \ref{eq:condition1} and \ref{eq:condition2}.
\end{thm}

\proof

First, one can easily observe that for any $a_1,a_2,a_3,$ such that $a_1+a_2+a_3 \leq 1$ and $a_1+a_2+a_3 \in \mathbb{R}_+ \cup \{0\}$ the Kraus operators proposed in \ref{eq:kraus2}-\ref{eq:kraus2con} constitute a completely positive map which is also strictly trace-preserving (i.e. it is a CPTP map) because the following equality holds
\begin{equation}
K_0 ^* (t;a_1,a_2,a_3,) K_0 (t;a_1,a_2,a_3)+ \sum_{i=1}^3 K_i ^* (t; a_i) K_i (t; a_i)  = \mathbb{I}_2.
\end{equation}
Therefore, the Kraus operators in \ref{eq:kraus2}-\ref{eq:kraus2con} constitute a quantum channel only if the parameters fulfill the condition \ref{eq:condition1}. One can notice it is an example of time-dependent \textit{Pauli channel}, which is often analyzed in the context of quantum information theory and quantum computing \cite{wudarski13}.

Consequently, $\rho(t)$ at any time can be computed from the dynamical map
\begin{equation}
 \rho (t) = K_0 (t;a_1,a_2,a_3,) \rho(0) K_0 ^* (t;a_1,a_2,a_3)+ \sum_{i=1}^3 K_i (t; a_i) \rho(0) K_i ^* (t; a_i)
\end{equation}
One can calculate the derivative of $\rho(t)$, which leads to the evolution equation in the Kossakowski-Lindblad form
\begin{equation}\label{eq:kossakowski2}
\frac{d \rho}{d t} = \gamma \left ( a_1 \sigma_1 \rho \sigma_1 + a_2 \sigma_2 \rho \sigma_2 + a_3 \sigma_3 \rho \sigma_3 - (a_1+a_2+a_3) \rho \right ).
\end{equation}
Applying to this equation the relation that connects the Cauchy product with the Kronecker product \cite{henderson81}, i.e.
\begin{equation}
vec(XYZ) = (Z^T \otimes X) vec Y,
\end{equation}
where it is assumed that the matrix product $XYZ$ is computable, one gets the explicit matrix form of the generator of evolution
\begin{equation}\label{eq:generator2}
\mathbb{L} = \gamma \left ( a_1 \sigma_1 \otimes \sigma_1 + a_2 \sigma_2^T \otimes \sigma_2 + a_3 \sigma_3 \otimes \sigma_3 - (a_1+a_2+a_3) \mathbb{I}_4 \right ).
\end{equation}

Now one can instantly notice that the parameter $\gamma$ multiplies the generator of evolution and, therefore, it does not change its structure. This observation explains why in the theorem \ref{The2} the family of Kraus operators was introduced as three-parametric. One can also re-introduce the parameters by substituting
\begin{equation}
\tilde{a_1} = \gamma a_1,\hspace{0.5cm} \tilde{a_2} = \gamma a_2  \hspace{0.5cm}and\hspace{0.5cm}  \tilde{a_3} = \gamma a_3,
\end{equation}
which gives the generator of evolution in the form:
\begin{equation}\label{eq:tildegen}
\mathbb{L} =  \tilde{a_1} \sigma_1 \otimes \sigma_1 + \tilde{a_2} \sigma_2^T \otimes \sigma_2 + \tilde{a_3} \sigma_3 \otimes \sigma_3 - (\tilde{a_1}+\tilde{a_2}+\tilde{a_3}) \mathbb{I}_4.
\end{equation}
Having the generator of evolution in the form \ref{eq:tildegen}, one can easily agree that in the analyzed case we deal with three parameters.

The matrix form of the operator $\mathbb{L}$ enables one to calculate the eigenvalues which depend on the parameters in the following way
\begin{equation}\label{eq:eigens}
\alpha_1 =  0, \alpha_2 = -2(a_1+a_2) \gamma,  \alpha_3 = -2(a_1+a_3) \gamma,  \alpha_4 = -2 (a_2+a_3) \gamma.
\end{equation}

Bearing in mind that the parameters have to fulfill \ref{eq:condition1}, one can notice that the spectrum of the generator of evolution consists of four different eigenvalues, i.e. $\alpha_1 \neq \alpha_2 \neq \alpha_3 \neq \alpha_4$, only if it will be additionally assumed that $a_1 \neq a_2  \neq a_3$, which means that the index of cyclicity of the geneator \ref{eq:generator2} is equal 1. This analysis proves that the generator of evolution \ref{eq:generator2} can be considered an optimal evolution model or, equivalently, the Kraus operators given in \ref{eq:kraus2}-\ref{eq:kraus2con} constitute a non-degenerate family.
\endproof

It this section it has been proved that there exists a three-parametric non-degenerate family of Kraus operators for 2-level systems. It means that with any set of parameters $(a_1,a_2,a_3)$ that fulfill the conditions \ref{eq:condition1} and \ref{eq:condition2} corresponds a generator of evolution $\mathbb{L}$ given by \ref{eq:generator2} which does not have any degenerate eigenvalues, i.e. its index of cyclicity is equal 1. According to the general results of the stroboscopic tomography in such a case there exists one observable the measurement of which performed at three different time instants provides enough data to reconstruct the initial state.

The desired observable, denoted by $Q$, has to fulfill the condition \cite{jam83,jam12}
\begin{equation}\label{eq:span}
Span\{\mathbb{I}_2, Q, \mathbb{L}^* [Q], (\mathbb{L}^*)^2 [Q] \} = B_* (\mathcal{H}),
\end{equation}
where $B_* (\mathcal{H})$ refers to the space of all Hermitian bounded linear operators in $\mathcal{H}$. This condition means that we require from the set $\{\mathbb{I}_2, Q, \mathbb{L}^* [Q], (\mathbb{L}^*)^2 [Q] \}$ to be complete. In the condition \ref{eq:span} by $\mathbb{L}^*$ one should understand the operator that governs the evolution of observables (in other words it is the dual operator or Heisenberg generator). Naturally, the Kossakowski-Lindblad equation for a density matrix $\dot{\rho} = \mathbb{L} [\rho]$ refers to the Schr{\"o}dinger picture of quantum mechanics, whereas in the Heisenbeg picture one would have an equation for an observable evolving in time, i.e. $\dot{Q} = \mathbb{L}^* [Q]$. The necessity to consider the Heisenberg picture arises straight from the basic considerations in the stroboscopic tomography. It was already mentioned in the introduction and to find more details one can browse a very well written paper \cite{jam12}. Therefore, in order to consider the condition \ref{eq:span} one needs to either transform the Kossakowski-Lindblad equation \ref{eq:kossakowski2} into the equation for evolving observables or one can multiply the Hermitian conjugate of the matrix form $\mathbb{L}$ ($4-$dimensional matrix) by vectorized observables, which shall be denoted by $\mathbb{L}^* vec Q$. We shall follow the other way of dealing with the condition \ref{eq:span} because the computation is faster.

Clearly for any $(a_1,a_2,a_3)$ that obey \ref{eq:condition1} and \ref{eq:condition2} there is an infinite number of observables that fulfill the condition from eq. \ref{eq:span}. If one wants to determine a general (parametric-dependent) structure of such observables, one can first notice that for $dim\mathcal{H} =2$ we consider observables matrices of the form
\begin{eqnarray}\label{eq:observable}
Q = \left ( \begin{array}{ll} A & C + iD \\ C - iD & B \end{array} \right ),  
\end{eqnarray}
where $A,B,C,D \in \mathbb{R}$ and $i$ denotes $\sqrt{-1}$. 

One can easily vectorize observable $Q$ to get
\begin{eqnarray}
vec Q = \left ( \begin{array}{l} A \\ C- iD \\ C+iD \\ B \end{array} \right ), 
\end{eqnarray}
which allows us to consider a $4-$dimensional matrix:
\begin{eqnarray}
M  = \left ( \begin{array}{llll} &  &   &   \\  &  &   &   \\ vec \mathbb{I}_2 & vec Q & \mathbb{L}^* vec Q  & (\mathbb{L}^* )^2 vec Q \\   &  &     &   \end{array} \right ).
\end{eqnarray}

The condition $\ref{eq:span}$ can be substituted by the condition that $det(M) \neq 0 $ due to the fact that the vectors $ vec \mathbb{I}_2, vec Q, \mathbb{L}^* vec Q, (\mathbb{L}^* )^2 vec Q$ have to be linearly independent. One can calculate that $det(M) \neq 0 $ if and only if 
\begin{equation}\label{eq:obcon}
A\neq B   \hspace{0.5cm}and\hspace{0.5cm}   C \neq 0  \hspace{0.5cm}and\hspace{0.5cm}  D\neq 0.
\end{equation}

Therefore, an observable of the structure \ref{eq:observable} such that its elements satisfy \ref{eq:obcon} fulfills the necessary condition for quantum tomography \ref{eq:span}. One can conclude that the constraints for an observable \ref{eq:obcon} are not particularly strict. Therefore, when thinking of future applications in experiments, there is a relatively wide choice of observables such that one of them is sufficient to perform quantum tomography on a quantum system with dynamics given by \ref{eq:kraus2}-\ref{eq:kraus2con} with conditions \ref{eq:condition1} and \ref{eq:condition2}. In order to reconstruct the initial density matrix one needs to perform the measurement of the observable $Q$ satisfying the conditions \ref{eq:obcon} at three different time instants, which will lead to obtaining the projections $\langle Q|\rho(0) \rangle$, $\langle \mathbb{L}^* [Q]|\rho(0) \rangle$ and $\langle (\mathbb{L}^*)^2 [Q]|\rho(0) \rangle$. The knowledge about these projections is sufficient for density matrix reconstruction.

Finally, in this section we may refer to the one-parametric non-degenerate family of Kraus operators $\{K_i (t; a) \}_{i=0}^{2}$ which was introduced in \cite{czerwin15}. It has the following form.
\begin{eqnarray}
 K_0 (t;a) = \sqrt{\frac{1+ 2 \kappa(t)}{3}} \mathbb{I},\hspace{0.75cm}  K_1(t;a) = \sqrt{\frac{a(1-\kappa(t))}{3}} \sigma_1,\\
K_2(t;a) = \sqrt{\frac{(2-a)(1-\kappa(t))}{3}} \sigma_2.
\end{eqnarray}

In \cite{czerwin15} it was proved that such a family is a one-parametric non-degenerate family of Kraus operators iff $ a \in \mathbb{R}$ and $a\in (0;2)$. One can easily notice that the family proposed in \cite{czerwin15} is a special case of the three-parametric non-degenerate family for $a_1 = \frac{a}{3}, a_2 = \frac{2-a}{3}, a_3 = 0$.

In this section it has been proved that there exists a non-degenerate family of Kraus operators that depends on three parameters $(a_1,a_2,a_3)$ which have to satisfy \ref{eq:condition1} and \ref{eq:condition2}. We have also proposed the general structure of the observable that allows to reconstruct such a state. One can agree that due to multi-parametric approach in case of both structure of the generator of evolution and the structure of the observables the analysis presented in this section possesses great potential applications in experiments. Finally,  one should bear in mind that the results introduced in this section are a generalization of the idea presented for the first time in \cite{czerwin15}.

\section{Six-parametric non-degenerate family of Kraus operatos for 3-level systems}\label{sec3}

Looking for a non-degenerate family of Kraus operators for 3-level systems we shall take into consideration the generators of $SU(3)$. By $\lambda_i$ for $i=1,..., 8$ we shall denote the set of the Gell-Mann matrices, which are a generalization of Pauli matrices.

We shall assume a family of Kraus operators in the following form
\begin{equation}\label{eq:kraus3}
 K_0(t;a_1,\cdots,a_8) = \sqrt{1-f(a_1,...,a_8)\left (1-\kappa(t)\right )}\mathbb{I}_3, \left \{ K_i (t;a_i) = \sqrt{a_i(1-\kappa(t))} \lambda_i \right \}_{i=1} ^8,
\end{equation}
where $a_i \in \mathbb{R}_+ \cup \{0\}$ are the parameters that influence the structure of the generator of evolution and $f$ is a certain function with condition $f(a_1,...,a_8)\leq 1$. The specific form of $f$ shall be determined later.

Due to the fact that the Kraus operators should constitute a quantum channel we obtain a relation
\begin{equation}
\sum_{i=1} ^8 K_i^* (t;a_i) K_i (t;a_i) \sim \mathbb{I}_3,
\end{equation}
which leads to two relations between parameters $a_1,...,a_8$
\begin{eqnarray}
a_7 = a_4+a_5-a_6,\label{eq:a7} \\
a_8 = a_1+a_2+a_3-a_4-a_5,\label{eq:a8}
\end{eqnarray}
which means that from the initial set only six parameters $a_1,...,a_6$ are linearly independent.

Taking into account \ref{eq:a7} and \ref{eq:a8}  it can be observed that 
\begin{equation}
\sum_{i=1} ^8 K_i^* (t;a_i) K_i (t;a_i) = \frac{2}{3} \left ( 2a_1+2a_2+2a_3+ a_4+a_5 \right ) (1-\kappa(t)) \mathbb{I}_3.
\end{equation}
Thus the explicit form of the $K_0 (t;a_1,...,a_8)$ can established
\begin{equation}
K_0(t;a_1,...,a_8) = \sqrt{1-\frac{2}{3} \left ( 2a_1+2a_2+2a_3+ a_4+a_5 \right ) (1-\kappa(t))} \mathbb{I}_3,
\end{equation}
which ensures that the Kraus operators from \ref{eq:kraus3} constitute a quantum channel.

The foregoing analysis allows us to propose the following theorem.
\begin{thm}\label{The4}
A six-parametric family of Kraus operators for 3-level systems given by
\begin{eqnarray}
 K_0(t;a_1,...,a_5) = \sqrt{1-\frac{2}{3} \left ( 2a_1+2a_2+2a_3+ a_4+a_5 \right ) (1-\kappa(t))} \mathbb{I}_3,\label{kraus001}\\
\left \{ K_i (t;a_i) = \sqrt{a_i(1-\kappa(t))} \lambda_i \right \}_{i=1} ^6,\\
 K_7(t;a_4,a_5,a_6) = \sqrt{(a_4+a_5-a_6)(1-\kappa(t))} \lambda_7 \\
K_8 = \sqrt{(a_1+a_2+a_3-a_4-a_5)(1-\kappa(t))} \lambda_8, \label{kraus002}
\end{eqnarray}
where $a_1,...,a_6$ are parameters that influence the structure of the generator of evolution, \\ can be considered a six-parametric non-degenerate family of Kraus operators only if the parameters $a_i$ fulfill the following conditions
\begin{eqnarray}
a_1, a_2,a_3,a_4,a_5,a_6, a_4+a_5-a_6,a_1+a_2+a_3-a_4-a_5 \in \mathbb{R}_+ \cup \{0\},\label{eq:1}\\
\frac{2}{3} \left ( 2a_1+2a_2+2a_3+ a_4+a_5 \right ) \leq 1,\label{eq:2} \\
a_1 \neq a_2 \neq a_3 \neq a_4 \neq a_5 \neq a_6 \neq a_4+a_5-a_6 \neq a_1+a_2+a_3-a_4-a_5.\label{eq:3}
\end{eqnarray}

\end{thm}
\proof

It can be observed that the conditions \ref{eq:1} and \ref{eq:2} are necessary for the Kraus operators \ref{kraus001}-\ref{kraus002} to constitute a quantum channel.

The generator of evolution for a 3-level system which is subject to decoherence given by \ref{kraus001}-\ref{kraus002} can be written as
\begin{equation}
 \mathbb{L} = a_1 \lambda_1 \otimes \lambda_1 + a_2 \lambda_2 ^T \otimes \lambda_2 +a_3 \lambda_3 \otimes \lambda_3 +a_4 \lambda_4 \otimes \lambda_4 +a_5 \lambda_5 ^T \otimes \lambda_5 + a_6 \lambda_6 \otimes \lambda_6 + 
\end{equation}
$$(a_4+a_5-a_6)\lambda_7^T \otimes \lambda_7+ (a_1+a_2+a_3-a_4-a_5) \lambda_8 \otimes \lambda_8-\frac{2}{3} \left ( 2a_1+2a_2+2a_3+ a_4+a_5 \right )\mathbb{I}_9.$$
And its eigenvalues take the forms
\begin{eqnarray}
\alpha_1 = 0,\\
\alpha_2 = -2a_1 -2a_2 - a_4 - a_5,\\
\alpha_3 = -2a_1 - 2a_3 - a_4 - a_5,\\
\alpha_4 = -2a_2 - 2a_3 - a_4 - a_5,\\
\alpha_5 = -2a_1 -2a_2 - 2a_3 + a_4 - a_5,\\
\alpha_6 = -2a_1 -2a_2 - 2a_3 - a_4 + a_5,\\
\alpha_7 = -3(a_4+a_5),\\
\alpha_8 = -2a_1 -2a_2 - 2a_3 + a_4 + a_5-2a_6,\\
\alpha_9 = -2a_1 -2a_2 - 2a_3 - a_4 - a_5+2a_6,
\end{eqnarray}

We want to prove that in the spectrum of the generator $\mathbb{L}$ there are no two eigenvalues equal to each other if and only if the conditions \ref{eq:3} are satisfied. We shall demonstrate that the conditions \ref{eq:3} imply that the spectrum of $\mathbb{L}$ contains no degenerate eigenvalues because the proof of  the other implication is obvious. An elaborate proof can be conducted by means of the proof contradiction. If one assumes that certain two eigenvalues of $\mathbb{L}$ are equal, e.g. $\alpha_6 = \alpha_8$, then one always gets a contradiction with the assumptions \ref{eq:1} - \ref{eq:3}. In the example $\alpha_6 = \alpha_8$ implies that $a_4 = a_6$, which disagrees with the assumption \ref{eq:3}. In the same way one could consider all pairs $\alpha_i, \alpha_j$ for$i \neq j$ and putting $\alpha_i = \alpha_j$ one would always get a contradiction with the assumption \ref{eq:3}.
\endproof

Therefore, one can conclude that the assumptions about the parameters $a_i$ given by \ref{eq:1} - \ref{eq:3} are sufficient to claim that the spectrum of the generator $\mathbb{L}$ consists of 9 different eigenvalues, which means that for any evolutions given by \ref{kraus001}-\ref{kraus002} with parameters satisfying \ref{eq:1} - \ref{eq:3} there exists one observable the measurement of which performed at 8 different time instants provides enough data to reconstruct the initial state (and consequently the trajectory). This results shows that the idea of parametric-dependent non-degenerate families of Kraus operators can be extended to 3-level quantum systems.

\section{Optimal evolution models for tomography in case of $dim \mathcal{H} = n$}\label{sec4}

In this section we present some thoughts concerning optimal evolution models for tomography in the general case when $dim \mathcal{H} = n$ and $dim B_* (\mathcal{H}) = n^2$. At this point we can revise the definition of matrix discriminant \cite{basu03}. Let us assume that the generator of evolution depends on a set of $k$ parameters $ A = \{a_1, \dots, a_k\}$ and the generator itself shall be denoted by $\mathbb{L}(A)$ (naturally, it is a matrix of the form given in \ref{eq:generatormatrix}). Moreover, the eigenvalues of the generator will be denoted by $\lambda_1, \dots, \lambda_{n^2}$. Then we can define the discriminant of $\mathbb{L}(A)$ as
\begin{equation}
D[\mathbb{L}(A)] = \prod_{i<j}^{n^2} (\lambda_i - \lambda_j)^2.
\end{equation}

One can instantly notice that the generator $\mathbb{L}(A)$ has no degenerate eigenvalues if and only if
\begin{equation}\label{eq:discon}
D[\mathbb{L}(A)] \neq 0.
\end{equation}
Therefore, the condition \ref{eq:discon} means that the index of cyclicity of the generator of evolution is equal 1, which ensures that there exists one observable the measurement of which performed at $n^2 - 1 $ different time instants provides sufficient data to reconstruct the initial density matrix $\rho(0)$.

In the general case one can write the necessary condition for optimal quantum tomography analogously to the equation \ref{eq:span}. One gets
\begin{equation}\label{eq:span2}
Span\{ \mathbb{I}_n, Q, \mathbb{L}^*(A) [Q], (\mathbb{L}^*(A))^2 [Q], \dots, (\mathbb{L}^*(A))^{n^2 -2} [Q] \} = B_* (\mathcal{H}).
\end{equation}

One can notice that the question whether the condition \ref{eq:span2} can be satisfied or not is connected with the notion of the minimal polynomial of $\mathbb{L}(A)$, which shall be refered to as $\mu (\lambda, \mathbb{L}(A))$. We know that $\mathbb{L}(A) \in B (\mathcal{H} \otimes \mathcal{H})$, which implies that on the basis of the Cayley-Hamilton theorem $(\mathbb{L}(A))^{n^2}$ linearly depends on $\mathbb{I}_n, \mathbb{L}(A), (\mathbb{L}(A)^*)^2, \dots,  (\mathbb{L}(A)^*)^{n^2-1}$ (the same theorem obviously applies to the dual operator). Moreover, one can observe that zero always belongs to the spectrum of $\mathbb{L}(A)$ due to the fact that for every generator of evolution there exists a stationary state. Thus, the degree of the minimal polynomial of $\mathbb{L}(A)$ has to fulfill the inequality $deg \mu (\lambda, \mathbb{L}(A)) \leq n^2 - 1$. Naturally, the degree of the minimal polynomial of $\mathbb{L}(A)$ is the same as in case of dual operator $\mathbb{L}^*(A)$. One can instantly notice that in order for the condition \ref{eq:span2} to be fulfilled the operators $\mathbb{L}^*(A), (\mathbb{L}^*(A))^2, \dots,  (\mathbb{L}^*(A))^{n^2-1}$ have to be linearly independent, which is equivalent to saying that the degree of the minimal polynomial of  $\mathbb{L}(A)$ is maximal and equal $n^2 - 1$.

All the considerations included in this section can be summarized in a theorem \ref{The4}.
\begin{thm}\label{The5}
Let $\mathbb{L}(A)$ denote a generator of evolution that depends on a set of parameters $A=\{a_1, \dots, a_k\}$ and which corresponds to a Hilbert space such that $dim \mathcal{H} = n$. Then the statement that the generator of evolution $\mathbb{L}(A)$ constitutes an optimal evolution model for quantum tomography can be equivalently expressed in three different ways. \\
1. The index of cyclicity of $\mathbb{L}$ is equal 1, i.e. $\max \limits_{\lambda \in \sigma (\mathbb{L})} \{ dim Ker (\mathbb{L} - \lambda \mathbb{I})\}=1$. \\
2. The degree of the minimal polynomial of $\mathbb{L}$ is equal $n^2 - 1$, i.e. $deg \mu (\lambda, \mathbb{L}(A)) = n^2 -1 $. \\
3. The discriminant of $\mathbb{L}(A)$ is non-zero, i.e. $D[\mathbb{L}(A)] \neq 0$.
\end{thm}

Here we introduced a few initial remarks concerning optimal evolution models for $n-$level quantum systems. The ideas outlined in this section shall be researched into in further articles.

\section{Summary}

In this article the idea of parametric-dependent non-degenerate families of Kraus operators has been extended in two ways. First, for 2-level quantum system a three-parametric family has been introduced, which seems significantly more general than the result presented in \cite{czerwin15}. Subsequently, the case of 3-level quantum systems has been considered, which led to the six-parametric non-degenerate family of Kraus operators. Both results mean that for an infinite number of generators of evolution there exists \textbf{one observable} (one physical quantity) the measurement of which performed at certain time instants provides enough data to reconstruct the initial state (and consequently the trajectory). Lastly, we introduced some general ideas concerning the optimal evolution models in case when $dim \mathcal{H} = n$.

The results introduced in this paper indicate great potential of the stroboscopic tomography for applications in experiments. The idea of non-degenerate family of Kraus operators means that in order to perform quantum tomography it is sufficient to prepare one experimental set-up and then repeat the same measurement certain number of times at different instants.

\section*{Acknowledgement}
This research has been supported by the grant No. DEC-2011/02/A/ST1/00208 of National Science Center of Poland.

\end{document}